\documentclass[pra,twocolumn,showpacs,preprintnumbers,amsmath,amssymb]{revtex4}

\usepackage{graphicx}
\usepackage{pst-plot, pstricks}
\usepackage{dcolumn}
\usepackage{bm}

\def\ket#1{{|#1\rangle}}
\def\s{{\{0,1\}^*}}
\def\hr{{\mathcal{H}}}
\def\C{{\mathbb{C}}}
\def\lmax{{l_{max}}}
\def\Tr{{\rm Tr}}
\def\N{{\mathbb{N}}}
\def\R{{\mathbb{R}}}
\newcommand{\nix}{{\rule{0pt}{2pt}}}
\newcommand{\qedd}{{\nix\nolinebreak\hfill\hfill\nolinebreak$\Box$}}
\newcommand{\qed}{{\qedd\par\medskip\noindent}}
\newcommand{\lineclear}{{\rule{0pt}{0pt}\nopagebreak\par\nopagebreak\noindent}}
\newtheorem{theorem}{Theorem}[section]

\newtheorem{lemma}[theorem]{Lemma}
\newtheorem{definition}[theorem]{Definition}

\newtheorem{example}[theorem]{Example}

\begin{document}

\title{Lossless quantum prefix compression for communication channels \\ that are always open}

\author{Markus M\"uller}
 \email{mueller@math.tu-berlin.de}
 \altaffiliation[also at ]{Institute of Mathematics, Berlin Institute of Technology (TU Berlin).}
\affiliation{%
Max Planck Institute for Mathematics in the Sciences\\
Inselstra\ss e 22, D-04103 Leipzig, Germany
}%
\author{Caroline Rogers}
 \email{caroline@dcs.warwick.ac.uk}
\author{Rajagopal Nagarajan}
 \email{biju@dcs.warwick.ac.uk}
\affiliation{
Department of Computer Science, University of Warwick\\
Coventry, CV47AL, United Kingdom
}%

\date{November 18, 2008}

\begin{abstract}
We describe a method for lossless quantum compression if the output of the information
source is not known. We compute the best possible compression rate, minimizing the expected base length
of the output quantum bit string (the base length of a quantum string is the maximal length
in the superposition). This complements work by Schumacher and Westmoreland who
calculated the corresponding rate for minimizing the output's average length.

Our compressed code words are prefix-free indeterminate-length quantum bit strings which can be concatenated in the case of
multiple sources. Therefore, we generalize the known theory of prefix-free quantum codes to the
case where strings have indeterminate length.
Moreover, we describe a communication model which allows the lossless transmission
of the compressed code words. The benefit of compression is then the reduction of transmission errors in the presence of noise.
\end{abstract}

\maketitle

\section{Introduction}
One of the main aims of information theory is to determine the most efficient
way to compress messages. The solution to this problem often reveals relations to entropy-like
quantities, as in Shannon's noiseless coding theorem~\cite{CoverThomas}, where the entropy of the information source
determines the best possible compression rate.

The situation in quantum information theory is quite similar. The most popular example is Schumacher's
noiseless coding theorem~\cite{Schumacher}, showing that the best possible compression rate in the quantum case is given
by von Neumann entropy. ``Compression'' here means that the number of qubits that have to be transmitted
to faithfully exchange a quantum state is minimized. This definition shows that the compression of quantum information
is automatically related to the problem of communication: once the compression is accomplished, then {\em how can
the compressed code words be transmitted to a receiver?}

This question addresses an important difficulty in the quantum situation which does not arise in classical
information theory: if a variable-length code is used for quantum compression, some code words will be shorter than
others. But this may result in code words which are in a superposition of different lengths --- how can those
code words be transmitted without disturbance?

This problem is one of several reasons why it was previously stated~\cite{classicalsidechannels,bos02,bra00,koa02,sch01}
that lossless compression of an ensemble $\mathcal{E}=\{p_i,|\psi_i\rangle\}$ of quantum states is in general {\em impossible}, if
the value $i$ of the state $\ket{\psi_i}$ to be compressed is unknown. A related objection~\cite{sch01} is that
{\em prefix-free codes} are also useless in the quantum situation: a prefix-free code word carries its own length
information. If it is transmitted over a channel, that length information must be read out to see when the transmission
is over and the channel can be closed. Again, if the code word is in a superposition of different lengths, this reading-out measurement
disturbs the code word.

In this paper, we show that the aforementioned problems do not appear if one uses a channel instead which
is {\em always open}. In this case, there is no need to decide when the transmission is finished. Even in the
case of such a channel, compression can be beneficial: it can help to reduce transmission errors.

To better understand the purpose of this paper, it makes sense to think about the compression of quantum
information as taking place in several steps:
\begin{itemize}
\item[Step 1.] First, the quantum state is {\em compressed}, typically yielding an output code which is in a superposition of
different lengths.
\item[Step 2.] Optional: The output code is {\em cut off} (projected) to get a determinate-length code (which introduces some loss).
\item[Step 3.] Finally, the code is {\em transmitted} over some quantum channel.
\end{itemize}
Actually, Schumacher and Westmoreland~\cite{sch01} give a method of this form for compression of quantum information using a prefix-free
quantum code. In fact, Step 1 in their setting is {\em lossless} --- it is a unitary and thus reversible
operation that minimizes the output's average length.

Then they show that a projection to the first $n\cdot (S+\delta)$ qubits does not disturb the message very much,
where $S$ is the source's entropy --- this is Step 2 in the scheme above, which introduces some (small) loss.
As the resulting code word consists of a classically known, determinate number of qubits, it is clear how to
transmit it over a channel (Step 3).

In this paper, we describe how Step 1 can be carried out losslessly if the task is to minimize the output's {\em base length}
instead of the average length (both length notions will be discussed in detail below, cf. Definition~\ref{DefDifferentLengths}),
and we compute the best possible compression rate in terms of an entropy-like quantity
(Theorem~\ref{MainTheorem} on page~\pageref{MainTheorem}) for the case of a single quantum information source.
We do this using {\em prefix-free quantum bit strings} such that code words can be concatenated in the
case of several sources.

For this reason, we advance the theory of prefix-free quantum strings, by generalizing the definition
and results of Schumacher and Westmoreland in a natural way.

Moreover, we explain that Step 3 is unproblematic if the channel in question is {\em always open},
even if Step 2 is dropped. All in all, this gives a {\em lossless} method of compression and transmission
of quantum information. The price we pay for it is that there is no way to see when the transmission is finished.
Yet, the benefit is that the probability of transmission errors can be reduced.

\section{Synopsis}
This paper is organized as follows:
\begin{itemize}
\item In Section~\ref{SecPreviousWork}, we give a brief description of previous work on lossless quantum compression.
In particular, we review the arguments why and in what way lossless quantum compression of  unknown states seems to be impossible.
We define indeterminate-length quantum bit strings (as used by several authors before) and give a physical interpretation.
\item In Section~\ref{SecCommunicationModel}, we give a communication model which describes a situation where lossless
quantum compression is possible and useful. In short, we explain the model of an ``always-open channel'' where neither
Alice nor Bob know when the transmission has finished, but both parties benefit from compression by reducing transmission errors.
\item Section~\ref{SecPrefixFree} contains a review of some results on prefix-free quantum bit strings,
generalizing work by Schumacher and Westmoreland~\cite{sch01}. We also give new results which
have useful interpretations in the framework of our compression scheme.
Moreover, we prove that the concatenation of prefix-free indeterminate-length quantum bit strings
can in principle be implemented physically.
\item Our main result is Theorem~\ref{MainTheorem} on page~\pageref{MainTheorem}. It states the optimal
rate for prefix-free compression of the unknown output of a single quantum information source, given the task to minimize the expected
base length.

To state the theorem, we define ``monotone entropy'' and ``sequential projections'' and discuss some properties
that simplify the computation of their actual numerical values.
\end{itemize}

\section{Previous Work}
\label{SecPreviousWork}
The aim of lossless quantum compression is to compress the unknown output $\ket{\psi_j}$ of an ensemble
$\mathcal{E}=\{(p_i,\ket{\psi_i})\}$ of quantum states using a variable-length
quantum code so that the original state $\ket{\psi_j}$ can always be retrieved
exactly and without error. When the $\ket{\psi_i}$'s are orthogonal,
this is equivalent to lossless classical compression. The challenge is therefore to encode $\mathcal{E}$
when the $\ket{\psi_i}$'s are non-orthogonal and the code words might have indeterminate lengths.

In this section, we first give a definition of quantum bit strings that consist of a superposition
of classical strings of different lengths. Then we describe previous work on how to use such quantum
strings for compression, and the difficulties that arise in such models. Finally, we outline a
physical interpretation of these indeterminate-length quantum strings.

\subsection{Indeterminate-Length Quantum Bit Strings}
The strategy of classical variable-length compression is to assign short code words $C(x)$
to frequent events $x$ (e.g. to frequent symbols in a text in some natural language),
while rare events are assigned the remaining long code words. Trying a similar approach
in quantum information theory naturally produces code words that are {\em superpositions
of classical strings of different lengths}.

For example, suppose we have two letters $A$ and $B$, and a classical code $C$ of the form
$C(A)=0$ and $C(B)=11$. If, as a first naive try, we extend this map unitarily to quantum states
spanned by $|A\rangle$ and $|B\rangle$, we get for example
\[
   C\left(\frac{\ket{A}+\ket{B}}{\sqrt{2}}\right)=\frac{\ket{0}+\ket{11}}{\sqrt{2}}
\]
which does not have a determinate length, since it is in a superposition of lengths $1$ and $2$.
It is called an {\em indeterminate-length quantum bit string}. Such strings can formally be defined as follows:
\begin{definition}[Quantum Bit String]
A quantum state $\ket{\psi}$ is a {\em quantum bit string} (or {\em qubit string}) if it is an element of the
Fock space (or {\em string space})
\[
   \hr_\s:=\bigoplus_{n=0}^\infty \left(\C^2\right)^{\otimes n},
\]
that is, if it can be expressed as a superposition of classical bit strings of the form
\[
   \ket{\psi} = \sum_{s\in\s}\alpha_s \ket{s}
\]
with $\alpha_s\in\C$ and $\sum_{s\in\s}|\alpha_s|^2=1$.
\end{definition}
For convenience, we will sometimes drop the normalization condition. Moreover,
it sometimes makes sense to call normal {\em mixed states}, i.e. density operators,
on $\hr_\s$ qubit strings, too. The reason is that the prefixes of pure qubit strings
can be mixed, which will be explained in detail below in Section~\ref{SecPrefixFree}.

Bostr\"{o}m and Felbinger~\cite{bos02} defined two ways to quantify the lengths
of indeterminate-length strings.
\begin{definition}[Length of Qubit Strings~\cite{bos02}]
\label{DefDifferentLengths}
\lineclear
The base length $L$ of an indeterminate-length string $\ket\psi=\sum_{s\in\s}\alpha_s\ket{s}$ is the length of the longest part of its superposition
\[
   L(\psi)=L\left(\sum_{s\in\s} \alpha_s \ket{s}\right) := \max_{\alpha_s\neq 0} \ell(s),
\]
or $\infty$ if the maximum does not exist.
This can also be written as $L(\psi)=\max\{\ell(s)\,\,|\,\,\langle s|\psi\rangle\neq 0\}$.
The average length $\bar\ell$
of an indeterminate-length quantum bit string is the expectation value of the length
\[
   \bar\ell(\psi)=\bar\ell\left(\sum_{s\in\s} \alpha_s \ket{s}\right) = \sum_{s\in\s} |\alpha_s|^2 \ell(s)
\]
which may as well be infinite. It can be written as $\bar\ell(\psi)=\langle\psi|\Lambda|\psi\rangle$, where $\Lambda$
is the length operator, defined by linear extension of
\[
   \Lambda\ket{s}=\ell(s)\ket{s}\qquad(s\in\s).
\]
Formally, $\Lambda$ is an unbounded self-adjoint operator, defined on a dense subspace of $\hr_\s$.
\end{definition}
If the length of a quantum string is observed, then $\bar\ell$ gives the expected length that is observed and $L$ gives the maximum length that can be observed. However, given an unknown indeterminate-length string $\ket{\psi}$,
neither its average length nor its base length can be measured without disturbing it.

\subsection{Can Indeterminate-Length Quantum Strings be Used for Coding?}

Various papers~\cite{classicalsidechannels,bos02,bra00,koa02,sch01} have described
problems in using indeterminate-length strings for lossless quantum data
compression. Braunstein {\it et al.}~\cite{bra00} pointed out three
difficulties of data compression with indeterminate-length strings.
The first is that if the indeterminate-length strings are unknown to
both the sender and the receiver, then it becomes impossible to synchronise
the different computational paths (taking different numbers of time steps)
that are performed on the strings.

The second difficulty is that if a
mixture of indeterminate-length strings is transmitted at a fixed
speed, then the recipient can never be sure when a message has
arrived and the strings can be decompressed. The third difficulty is
that if the data compression is performed by a read/write head (like a
Turing machine), then after the data compression, the head location
of the sender is entangled with the ``lengths" of the indeterminate-length
string which represents the compressed data.

Koashi and Nobuyuki~\cite{koa02} argued that it is impossible to faithfully encode a mixture of
non-orthogonal quantum states if the particular output states of the quantum information source {\em are not known}.
They modelled lossless data compression as taking
place in a register of $N$ qubits. A compressed state
in the register would be an unknown indeterminate-length quantum string with base length $L$,
in which case, only the remaining $N-L$ qubits would be usable by other applications without disturbing
the compressed state. However the base length $L$ is not an observable, thus the other applications
cannot determine how many qubits are available. Thus the remaining $N-L$ qubits are not available
for other applications to use, unless there is some a priori knowledge about $L$ for some reason.

Schumacher and Westmoreland~\cite{sch01} showed that lossless quantum compression
cannot be carried out by a unitary operation in a simple model of communication.
They envisaged that indeterminate-length quantum strings would be padded with zeros to
create determinate length strings (we explain this in more detail below in Subsection~\ref{SubsecProperties}).
They modelled the data compression as taking place
between two parties Alice and Bob in which Alice sends Bob only the
original strings (with the zero-padding removed) leaving Alice with
a number of zeros depending on the length of the string she sent. If
she sends Bob an indeterminate-length string, then after the
transmission, Alice and Bob are entangled by the number of zeros that
are left on Alice's register.

Bostr\"{o}m and Felbinger~\cite{bos02} argued that it is not useful to consider quantum generalizations of
classical prefix-free codes:
classical prefix-free strings carry their own length information,
but the length information in an indeterminate-length quantum string cannot be observed without
disturbing the string. Their solution to this problem is to use a classical side channel to inform
the receiver where to separate the code words.

Ahlswede and Cai~\cite{classicalsidechannels} followed the same idea by sending the length information
over a classical side channel. Compared to Ref.~\cite{bos02}, they improved the compression rate
by giving a more efficient way to use the side channel, and they characterized the optimal compression
rate in this setting. We describe both approaches in more detail below in Subsection~\ref{SubsecClassicalSideChannels}.

However, in both cases, the use of the classical side channel requires that the sender (Alice)
knows the output of the quantum information source (at least partially), and thus the length of the compressed code word.
This is in contrast to the situation examined in this paper.

\subsection{Schumacher and Westmoreland's Prefix-Free Average Length Compression}
\label{SubsecProperties}

Schumacher and Westmoreland~\cite{sch01} investigated the general properties of
indeterminate-length strings. An indeterminate-length string can be
padded with zeros such that it consists of a determinate number of qubits.
\begin{definition}[Zero-extended form]
\label{DefZeroExtendedForm}
\lineclear
If $\ket{\psi}=\sum_{\ell(s)\leq l_{\max}} \alpha_s \ket{s}$ is a quantum string in
a register of $\lmax$ qubits, then its zero-extended form is
\[
   \ket{\psi_{\rm{zef}}} =\sum_{\ell(s)\leq l_{max}} \alpha_s \ket{s 0^{\otimes \lmax - \ell(s)}}.
\]
This string has a determinate length of $l_{max}$ qubits.
\end{definition}
Given a sequence of $N$ strings, it is useful to be able to ``concatenate'' them so that
the strings are packed together at the beginning of the string and the zero-padding
all lies at the end of the sequence. Schumacher and Westmoreland~\cite{sch01} call this the
``condensation operation''.
\begin{definition}[Condensable strings~\cite{sch01}]
\lineclear
A code is {\em condensable} if for every $N$, there exists a unitary operation
$U$ such that
\[
   U(\ket{\psi^{1}_{\rm{zef}}}\otimes \ldots \otimes \ket{\psi^{N}_{\rm{zef}}}) = (\ket{\psi^{1}}\otimes \ldots \otimes \ket{\psi^{N}})_{\rm{zef}}
\]
for all the code words $\ket{\psi^i}$ which are ``length eigenvectors'', i.e. if each $\ket{\psi^i}$ contains in its
superposition only classical words of some fixed length.
\end{definition}
For example, if $\ket{\psi^1}=\ket{0}$, $\ket{\psi^2}=\ket{10}$
and $\ket{\psi^3}=\ket{111}$, then
the condensation operation $U$ is
\begin{eqnarray*}
U(\ket{\psi^{1}_{\rm{zef}}}\otimes \ket{\psi^{2}_{\rm{zef}}}\otimes \ket{\psi^{3}_{\rm{zef}}})
&=& U(\ket{000}\otimes \ket{100}\otimes \ket{111}) \\
&=& \ket{0}\otimes\ket{10}\otimes\ket{111}\otimes\ket{000} \\
&=& (\ket{0}\otimes\ket{10}\otimes\ket{111})_{\rm{zef}}.
\end{eqnarray*}

Superpositions of classical prefix-free strings are condensable.
More generally, Schumacher and Westmoreland gave a definition of prefix-free quantum strings and showed that
they are condensable.
According to their definition, two strings are prefix-free if when the additional qubits in the longer string are traced out,
the resulting prefixes are orthogonal.
\begin{definition}[Zero-padded prefix-freedom~\cite{sch01}]
\label{schumacherprefixfree}
Suppose $\ket{\psi}$ and $\ket{\varphi}$ are quantum strings with $n:=L(\ket{\psi})<L(\ket{\varphi})$
and that they are in a register of $\lmax$ qubits.
The first $n$ qubits of $\ket{\varphi_{\rm{zef}}}$ may
be in a mixed state, described by the density operator
\[
   \rho^{1 \ldots n} = \Tr_{n+1, \ldots, \lmax}(|\varphi_{\rm zef}\rangle\langle\varphi_{\rm zef}|).
\]
The strings $\ket{\psi}$ and $\ket{\varphi}$ are prefix-free if
\[
   \langle\psi| \rho^{1 \ldots n} \ket{\psi} =0.
\]
\end{definition}
This definition assumes that the two strings have determinate length. Two of the authors defined
prefix-free strings more generally such that they can be supported on subspaces which are spanned
by indeterminate-length quantum bit strings~\cite{markusprefix}. We give a review of this more general definition
in Section~\ref{SecPrefixFree}, which in fact contains Definition~\ref{schumacherprefixfree} as a theorem
(Lemma~\ref{LemDistinguish}).

Given many copies $\mathcal{E}^{\otimes n}$ of a quantum information source ${\mathcal{E}}$, Schumacher and Westmoreland further showed how to
use prefix-free quantum bit strings for lossless compression (this corresponds to ``Step 1'' of the compression process as described
in the Introduction) using appropriate unitary operations. The indeterminate-length output is then
projected onto the first $n(S(\mathcal{E})+\delta)$ qubits (``Step 2''), where $S$ is von Neumann entropy. This projection (or partial trace)
introduces only a small error which vanishes in the asymptotic case $n\to\infty$.

This can be seen as follows:
Let $\rho$ be the density operator corresponding to $\mathcal{E}$, with spectral decomposition
\[
   \rho = \sum_i p_i |i\rangle\langle i|.
\]
Then $\mathcal{E}$ can be compressed by encoding each $\ket{i}$ as a prefix-free string
of length $\lceil-\log(p_i)\rceil$ with zero-padding.
$\rho^{\otimes n}$ can be compressed in the same
fashion, by encoding every factor individually and applying the condensation operation to the resulting code words.
Every string $\ket{\psi}$ in the typical subspace of $\rho^{\otimes n}$ has probability
$\langle\psi| \rho \ket{\psi}$
arbitrarily close to $2^{-nS(\rho)}$ as $n$ grows large. Thus, vector states in the typical subspace of $\rho$
are encoded in a classical manner as strings of length arbitrarily close to $nS(\rho)$. The image of the projection on the first
$n(S(\rho)+\delta)$ qubits thus contains this typical subspace.
As with overwhelming probability, the output is very close to this subspace,
it can afterwards be decoded with high (but not perfect) fidelity.

Hence this compression scheme consists of two parts as already mentioned in the Introduction: in a first step,
the quantum message is compressed losslessly, in the sense that the output has minimal {\em average length}
of about $n\cdot S(\mathcal{E})$. In a second part, some ``cut-off'' takes place, introducing some small error,
but preparing the output to be transmitted over conventional channels by transforming it to fixed length~\cite{chu00}.

One of the results of this paper is to show how the first step can be accomplished to minimize
the expected {\em base length} (in the case of a single source), thus complementing the work by Schumacher and Westmoreland.

\subsection{Compression with Classical Side Channels}
\label{SubsecClassicalSideChannels}
Bostr\"{o}m and Felbinger~\cite{bos02} gave a scheme for lossless
quantum compression of {\em known} ensemble outputs using classical side channels. If $\mathcal{E}=\{p_i,|\psi_i\rangle\}$ is
the mixture to be compressed, then they assume that the value of $i$
is known to the compressor, Alice. If she encodes $\mathcal{E}$ using a
unitary operation $C$, then she sends the base length of the
compressed string to Bob, the decompressor, through a classical side
channel. She then sends $L(C(\ket{\psi_i}))$ qubits of
$\ket{\psi_i}$'s zero-extended form to Bob. Since the length of the
encoded string is encoded classically, it is not necessary to use a prefix-free
code to encode the quantum part --- thus $C$ is unitary but not necessarily a condensation operation.

Ahlswede and Cai~\cite{classicalsidechannels} studied quantum data compression
with classical side channels in more detail. 
They found an expression
for the number of qubits that are sent through the quantum channel in Bostr\"{o}m and Felbinger's 
lossless quantum compression scheme~\cite{bos02}.

Moreover, they showed by using counterexamples that the optimal rate of compression $R$ of a one-one code
cannot be achieved by a greedy algorithm.
However, the main goal of Ahlswede and Cai
was to find a more efficient way to use the classical side channel
than just to report the base lengths. They showed that the quantum part
could be compressed further than in the scheme set out by Bostr\"{o}m and
Felbinger.

The basis for their scheme is as follows.
If $\mathcal{E}$ is the mixture to be compressed, and if there exists
some small subspace $X$ such that several states $\ket{\psi_i}$ lie exactly
within $X$, then
this fact can be reported through the
classical side channel. Thus the amount of quantum information that must be sent
through the quantum channel is reduced.
They gave an expression for the optimal rate of compression in their scheme 
of an ensemble $\mathcal{E}=\{p_i,|\psi_i\rangle\}$ when the states $\ket{\psi_i}$ are
linearly independent (but not necessarily orthogonal).

Compression with classical side channels
has been studied in more detail for lossy compression~\cite{lossycompwithclassical}.
Hayashi and Keiji~\cite{Hayashi} investigated variable- (but not indeterminate-) length universal compression.

Rallan and Vedral~\cite{ral02} gave another scheme for lossless
quantum compression with classical side channels which does not use
zero-extended forms. They envisaged that the compressed state would
be represented by photons --- thus using a tertiary alphabet
$\{\ket{0},\ket{1},\ket{\#}\}$, where $\ket{\#}$ denotes
the absence of a photon and marks the end of the string. They
assumed that Alice has $n$ copies of an ensemble $\mathcal{E}$ which
she would like to send to Bob. In this scheme, Alice only sends Bob
the value of $n$ through the classical channel. This
scheme has a nice physical interpretation.

\subsection{Physical Interpretation of Indeterminate-Length Strings}
\label{physical}

Bostr\"{o}m and Felbinger~\cite{bos02} pointed out that variable-length
quantum strings can be realised in a quantum system whose particle
number is not conserved. Rallan and Vedral~\cite{ral02} described in
detail an example system where the average length of a string can be
interpreted as its energy. 

A Hilbert space $H^{\otimes n}$ can be
realised by a sequence of photons $\ket{\phi_1}\otimes \ldots
\otimes \ket{\phi_n}$ in which $\ket{\phi_i}$ represents exactly one
photon with frequency $\omega_i$. The value of the qubit
$\ket{\phi_i}$ is realised by the polarisation of its photon, either
horizontal $\ket{0}$ or vertical $\ket{1}$. The absence of a photon at a particular frequency can
be represented by $\ket{\#}$ which is orthogonal to $\ket{0}$ and
$\ket{1}$. Indeterminate-length strings are obtained by allowing the
number of photons to exist in superposition and ordering the photons by their frequencies.
The first $\ket{\#}$ (which can be in a superposition of positions) is used to mark the end of the string.

The frequency of each
photon $\ket{\phi_i}$ is chosen to be approximately equal so that $\omega_i \approx
\omega$ for some value $\omega$. The energy in a superposition of photons
is the average energy required to either create or destroy that
superposition ($\hbar \omega$ per photon of frequency $\omega$ where $\hbar$
is Planck's constant).
Thus the energy of an indeterminate-length string of photons
$\ket{\phi}$ is proportional to its average length and is given by (approximately)
$\hbar \omega \bar\ell(\ket{\phi})$.

On the other hand, the base length of $\ket{\phi}$ represents the number of photons at different frequencies
that are used to describe $\ket{\phi}$. Thus the base length of $\ket{\phi}$ is the size of the system required to
carry the state $\ket{\phi}$.

\section{Communication Model for Lossless Quantum Compression}
\label{SecCommunicationModel}
Now we describe a model of a communication channel where lossless quantum compression of
unknown mixtures is possible and useful.

The main argument why lossless quantum compression of unknown states seems to be impossible
is that it is impossible to determine how many qubits to transmit when the message
is in a superposition of different lengths.
If Alice has an unknown indeterminate-length qubit string,
how can she find out when the transmission is finished and the channel can be closed?
To avoid this problem, we look at {\em always-open channels}.

\begin{figure}[!hbt]
\begin{center}
\includegraphics[angle=0, width=8.8cm]{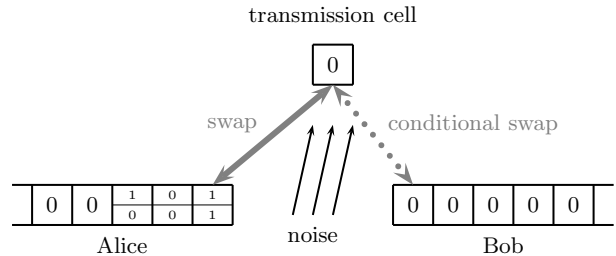}
\caption{Schematic of an always-open channel as described in Section~\ref{SecCommunicationModel}.}
\label{communication}
\end{center}
\end{figure}

A model of an always-open channel~\cite{bowen,nielsen} is shown in Fig.~\ref{communication}.
Suppose Alice wants to send Bob a single code word of a quantum prefix code, i.e.
an indeterminate-length qubit string $|\psi\rangle$ which is an element of a prefix-free
subspace $\hr$ of $\hr_\s$ that Alice and Bob have agreed upon in advance. (This single code word might itself be a concatenation of
several prefix-free code words.)
As we shall see later in Theorem~\ref{MainTheorem}, we may assume that $\hr$ is spanned by
the classical code words of a classical prefix code as in Subsection~\ref{SubsecProperties} above.

The main part of the channel is a transmission cell which carries exactly one qubit.
Initially, this qubit is set to zero, and so are all the qubits in Bob's memory.
Moreover, Alice's memory contains a zero-padded form of her message string, as
described in Definition~\ref{DefZeroExtendedForm}.

Now we describe the communication protocol --- for each step, we describe what Alice and Bob
are doing in the case of classical bits (i.e. in the case that the message qubit string $|\psi\rangle$
is just a classical string $|s\rangle$ out of the classical prefix-free orthonormal basis
of $\hr$), and we assume that the resulting operation
is linearly extended to a unitary operation on the corresponding quantum system. The unitarity
of the operations at Alice's and Bob's side is then assured by the reversibility of the corresponding classical operations.

At step $i$ of the transmission, Alice swaps the $i$-th qubit of her padded message string with the content
of the transmission cell. Afterwards, Bob checks if the $i-1$ qubits he has received previously form a valid
code word or not. (Due to prefix-freedom, if the answer is ``yes'', then the transmission must be over ---
cf. also Definition~\ref{schumacherprefixfree} and Lemma~\ref{LemDistinguish}). If the answer is ``no'', he swaps
the $i$-th qubit of his memory (which is just a zero) with the content of the transmission cell; otherwise,
he does not do anything. That is,
Bob applies a conditional swap, where the condition is that the transmission is not yet finished.

This way, the message qubit string is transmitted qubit by qubit. In the end, Alice ends up
with a memory full of zeroes, while the transmission cell contains a zero as well, and Bob's memory
carries the zero-padded message string. Hence the entanglement problem described by Schumacher and Westmoreland~\cite{sch01} is avoided,
and the message qubit string is transmitted reversibly and unitarily from Alice to Bob.

But what is the advantage of compression for such a communication channel, if that channel can never be switched off by Alice or Bob?
It cannot be used to save transmission time (considered as a resource), because both parties never know if the
transmission is already finished or not (unless some predefined maximal transmission time $t_{max}$ has
passed). However, quantum compression can have other advantages: for example, suppose the transmission cell is
subject to noise during the transmission. That is, the transmission of every single qubit
has an inherent error probability. In this case, Alice can minimize transmission errors by compressing
her quantum messages before sending them.

To be more exact, as soon as the code word has been fully transmitted (i.e. at a time step corresponding
to the message's base length), Bob stops to access the transmission cell. Thus, any noise that affects
the cell from that point on will not disturb the communication any more, because the channel to Bob
is effectively closed. Thus, minimizing the number of qubits to be transmitted reduces he probability
of transmission errors, even though neither Alice nor Bob know the number of transmitted qubits.

It is clear that the optimal compression method depends on the kind of noise that the system is exposed to.
Obviously, in the case that each transmitted qubit is independently subject to the same kind of perturbation,
then Schumacher and Westmoreland's average length compression method optimally minimizes transmission errors.
But there are other conceivable scenarios: for example, we might have several channels at once that are subject
to the same kind of noise, or time-dependent noise that grows with the number of qubits. In this case, it is not
so clear any more what the best method of compression is.

In this paper, we compute the best possible compression method and the rate for minimizing the code's expected
base length. Although we do not currently know of a natural noise model where the expected base length
determines the error probability, it seems likely that there are indeed natural situations
where this kind of compression is superior to average length compression --- for example, models
like those mentioned in the last paragraph where ``later'' qubits are subject to larger errors than ``earlier'' ones.

\section{Prefix-Free Quantum Bit Strings}
\label{SecPrefixFree}
Schumacher and Westmoreland~\cite{sch01}
defined prefix-free quantum strings in terms of their zero-extended forms using the partial trace, see Definition~\ref{schumacherprefixfree} above.
In Ref.~\cite{markusprefix}, two of the authors have given another way to define prefix-free quantum strings
which is more general and a more direct generalization of the classical definition.
It can be shown to contain the definition by Schumacher and Westmoreland as a special case.
In this section, we briefly review the definition and basic results on prefix-free quantum strings.

The notion of the prefix of a classical string is closely related to the concatenation operation $\circ$.
Thus, before we define prefix-free quantum strings,
we first explain how to concatenate quantum bit strings. If $\ket{\psi}\in\hr_\s$ is any quantum
bit string, and $s\in\s$ is a classical bit strings, then we can define $\ket{\psi}\circ s$
by linear extension of the classical concatenation: Expand $\ket{\psi}=\sum_{x\in \s} \alpha_x \ket{x}$, and define
\[
   |\psi\circ s\rangle:=\ket{\psi}\circ s:=\sum_{x\in s} \alpha_x \ket{x\circ s}.
\]
Moreover, if $|\varphi\rangle=\sum_{t\in\s}\beta_t |t\rangle$ is another qubit string with finite base length,
then we set
\[
   |\psi\circ \varphi\rangle:=|\psi\rangle\circ |\varphi\rangle:=\sum_{t\in\s}\beta_t |\psi\circ t\rangle.
\]
This concatenation operation on the quantum strings is related to the tensor product:
If $\ket{\psi}$ is a length eigenstate (i.e. an eigenvector of the length operator $\Lambda$),
then $\ket{\psi}\circ\ket{\varphi}=\ket{\psi}\otimes\ket{\varphi}$. However, if $\ket{\psi}$ is not
a length eigenstate, then the concatenation operation is not always an isometry and thus not
always physically meaningful~\cite{markusprefix}.

We can now define prefix-free sets of quantum strings (e.g. prefix-free subspaces of the string space $\hr_\s$)
by direct generalization of the classical definition. Although there are several a priori possible generalizations,
they all turn out to be equivalent (for a proof see Ref.~\cite{markusprefix}). To state them, we use the symbol
$\lambda$ for the empty string of length zero.

\begin{definition}[Prefix-Free Sets of Qubit Strings]
\label{DefPrefixFree}
A set $M\subset \hr_\s$ of qubit strings is called {\em prefix-free}, if one of the four
following equivalent conditions holds:
\begin{itemize}
\item[(1)] For every $|\varphi\rangle,|\psi\rangle\in M$ and classical string $s\in\s\setminus\{\lambda\}$, it holds
$\langle \varphi|\psi\circ s\rangle=0$.
\item[(2)] For every $|\varphi\rangle,|\psi\rangle\in M$ and qubit string $|\chi\rangle\perp |\lambda\rangle$, it holds
$\langle \varphi|\psi\circ\chi\rangle=0$.
\item[(3)] For every $|\varphi\rangle,|\psi\rangle\in M$ and classical strings $s,t\in\s$ with $s\neq t$, it holds
$\langle \varphi \circ t|\psi\circ s\rangle=0$.
\item[(4)] For every $|\varphi\rangle,|\psi\rangle\in M$ and qubit strings $|\chi\rangle,|\tau\rangle\in\hr_\s$ with $|\chi\rangle
\perp|\tau\rangle$, it holds
$\langle \varphi\circ\tau|\psi\circ\chi\rangle=0$.
\end{itemize}
\end{definition}

The relevant case for quantum compression is that $M$ is itself a closed subspace of string space $\hr_\s$.
To prove prefix-freedom of such a subspace, it is sufficient to prove this property for an arbitrary
orthonormal basis~\cite{markusprefix}:

\begin{lemma}
\label{LemPrefixFreeHR}
A subspace $\hr\subset \hr_{\{0,1\}^*}$ is prefix-free if and only if it has a prefix-free orthonormal basis.
In this case, {\em every} orthonormal basis of $\hr$ is prefix-free.
\end{lemma}

\begin{example}\label{ExStrangeHR}
The following subspace $\hr\subset\hr_\s$ is prefix-free:
\[
   \hr:={\rm span}\left\{
      \frac 1 {\sqrt{2}}\left(|1\rangle+|01\rangle\right), \frac 1 {\sqrt{2}} \left(|10\rangle-|010\rangle\right)
   \right\}
\]
It is easily checked that condition (1) from Definition~\ref{DefPrefixFree} above is satisfied for the two orthonormal basis vectors.
\end{example}

Similarly as in the classical case, closed prefix-free subspaces obey a Kraft inequality~\cite{markusprefix}:
\begin{lemma}[Quantum Kraft Inequality]
\label{LemKraft}
Let $\{|e_i\rangle\}_{i\in I}\subset\hr_\s$ be a prefix-free orthonormal system, spanning a closed subspace $\hr\subset\hr_\s$.
Then, it holds
\[
   \sum_{i\in I} 2^{-L(e_i)} \leq \sum_{i\in I} 2^{-\bar\ell(e_i)} \leq \Tr\left(2^{-\Lambda} \mathbb{P}(\hr)\right)\leq 1,
\]
where $\mathbb{P}(\hr)$ denotes the orthogonal projector onto $\hr$.
Equality holds for the left three terms if and only if every $|e_i\rangle$ is a length eigenvector.
\end{lemma}

Prefix-free subspaces have a remarkable property: every basis vector of length $n$ can be distinguished with
certainty from every other (even longer) basis vector by measuring the first $n$ qubits only. Unfortunately,
this is only true in general for orthonormal bases of length eigenvectors~\cite{markusprefix}.
\begin{lemma}
\label{LemDistinguish}
An orthonormal system $M\subset\hr_\s$ which consists entirely of length eigenvectors is prefix-free if and only if
for every $|\varphi\rangle,|\psi\rangle\in M$ with $|\varphi\rangle\neq |\psi\rangle$, it holds
\begin{equation}
   \langle \psi | \varphi^{\ell(\psi)} |\psi\rangle=0,
   \label{EqDistinguish}
\end{equation}
where $\varphi^n$ denotes the restriction of the quantum state $|\varphi\rangle\langle\varphi|$ to the first $n:=\ell(\psi)$ qubits.
\end{lemma}
This lemma shows that if the subspace contains an orthonormal basis of length eigenvectors,
our definition of prefix-freedom is equivalent to the definition by Schumacher and Westmoreland~\cite{sch01}.

We have only collected the basic facts about prefix-free quantum bit strings that are relevant
for lossless quantum compression.
For more details, we refer the reader to Refs.~\cite{sch01} and \cite{markusprefix}.

In general, the concatenation operation does not preserve the norm of vectors from $\hr_\s$, i.e.
it is not an isometry and hence not physically meaningful. However, we shall now prove that
concatenation can be implemented in principle on a quantum computer (i.e. it is an isometry) if one restricts to prefix-free Hilbert spaces:

\begin{theorem}[Isometry of Concatenation]
\label{TheIsoConcat}
If $\{\ket{\varphi_1},\ket{\varphi_2}\}\subset\hr_\s$ is a prefix-free set, and $\ket{\psi_1},\ket{\psi_2}\in\hr_\s$, then
\[
   \langle \varphi_1\circ \psi_1|\varphi_2\circ\psi_2\rangle=\langle \varphi_1|\varphi_2\rangle\langle \psi_1|\psi_2\rangle.
\]
Consequently, if $\hr\subset\hr_\s$ is a closed prefix-free subspace, then there exists a unique isometry
$U_\circ:\hr\otimes\hr_\s\to\hr_\s$ such that $U_\circ|\varphi\rangle\otimes|\psi\rangle=|\varphi\circ\psi\rangle$
for every $\ket\varphi\in\hr$ and $\ket\psi\in\hr_\s$.
\end{theorem}
Note that in the special case that $\hr$ is spanned by length eigenvectors, the map $U_\circ$ corresponds to
the ``simple condensation operation'' as defined by Schumacher and Westmoreland~\cite{sch01}.

{\bf Proof.} It is easy to check that for every pair of qubit strings $\ket{\varphi_1},\ket{\varphi_2}\in\hr_\s$ and $s\in\s$,
we have $\langle \varphi_1\circ s|\varphi_2\circ s\rangle=\langle \varphi_1|\varphi_2\rangle$.
Now suppose that additionally $\Phi:=\{\ket{\varphi_1},\ket{\varphi_2}\}$ is a prefix-free set, and $\ket\psi\in\hr_\s$
is an arbitrary qubit string. Expanding $\ket\psi=\sum_{s\in\s} \gamma_s \ket s$, we have
\begin{eqnarray*}
   \langle \varphi_1\circ\psi|\varphi_2\circ\psi\rangle&=&\sum_{s,t\in\s} \bar \gamma_s \gamma_t \langle
   \varphi_1\circ s|\varphi_2\circ t\rangle\\
   &\stackrel{(*)}=& \sum_{s\in\s} \bar\gamma_s \gamma_s \langle\varphi_1\circ s|\varphi_2\circ s\rangle \\
   &=&\langle \varphi_1|\varphi_2\rangle\sum_{s\in\s} |\gamma_s|^2 =\langle \varphi_1|\varphi_2\rangle\langle\psi|\psi\rangle.
\end{eqnarray*}
In $(*)$, we have used the fact that $\Phi$ is prefix-free, and so $\langle\varphi_1\circ s|\varphi_2\circ t\rangle=0$
if $s\neq t$. Finally, if $\ket{\psi_1},\ket{\psi_2}\in\hr_\s$ are arbitrary qubit strings, then choose an arbitrary
orthonormal basis $\{\ket{e_i}\}_{i\in\N}$ of $\hr_\s$ such that $\ket{\psi_1}=\lambda\ket{e_1}$ with $\lambda\in\R$, and expand $\ket{\psi_2}$
as $\ket{\psi_2}=\sum_{i\in\N} \alpha_i \ket{e_i}$. It follows
\begin{eqnarray*}
   \langle \varphi_1\circ\psi_1|\varphi_2\circ\psi_2\rangle&=&\sum_{i\in\N} \alpha_i \langle \varphi_1\circ\psi_1 |\varphi_2\circ e_i\rangle\\
   &\stackrel{(**)}=& \alpha_1 \langle\varphi_1\circ \psi_1|\varphi_2\circ e_1\rangle\\
   &=&\alpha_1\lambda\langle\varphi_1|\varphi_2\rangle\underbrace{\langle e_1|e_1\rangle}_{=1}\\
   &=&\langle\psi_1|\psi_2\rangle\langle\varphi_1|\varphi_2\rangle.
\end{eqnarray*}
In $(**)$, we have again used the fact that $\Phi$ is prefix-free, and consequently $\langle\varphi_1\circ\psi_1|\varphi_2\circ e_i\rangle
=0$ for $i\geq 2$, since $\ket{\psi_1}\perp \ket{e_i}$.
\qed

We show now that the base length of a concatenation of two qubit strings is the sum of the individual base lengths.
Note that this is in general not true for average length $\bar\ell$: for example, if
$|\psi\rangle=\frac 1{\sqrt{2}}(|1\rangle+|01\rangle)$ and $|\varphi\rangle=\frac 1 {\sqrt{2}}(|10\rangle-|010\rangle)$
are two vectors from the prefix-free Hilbert space $\hr$ in Example~\ref{ExStrangeHR}, and if
$|\chi\rangle:=\frac 1 {\sqrt{2}}(|\psi\rangle+|\varphi\rangle)$, then it is easy to check that
$\frac {19} 4 =\bar\ell(\chi\circ\varphi)>\bar\ell(\chi)+\bar\ell(\varphi)=2+\frac 5 2$.
\begin{lemma}[Additivity of Base Length]
\label{LemAdditivity}
\lineclear
If $\ket{\varphi},\ket\psi\in\hr_\s$ are qubit strings with finite base lengths,
i.e. $L(\varphi)<\infty$ and $L(\psi)<\infty$, then $L(\varphi\circ\psi)=L(\varphi)+L(\psi)$.
\end{lemma}
{\bf Proof.} For every $\ket\varphi\in\hr_\s$, define $S(\varphi):=\{s\in\s\,\,|\,\,\langle s|\varphi\rangle\neq 0\}$.
It follows that $L(\varphi)=\max\{\ell(s)\,\,|\,\,s\in S(\varphi)\}$. If we expand $\ket\varphi=:\sum_{s\in\s} \alpha_s\ket{s}$
and $\ket\psi=:\sum_{t\in\s}\beta_t\ket{t}$, then
\[
   |\varphi\circ\psi\rangle=\sum_{s,t} \alpha_s \beta_t |s\circ t\rangle.
\]
It follows that
\[
   S(\varphi\circ\psi)\subseteq S(\varphi)\circ S(\psi)
   :=\{s\circ t\,\,|\,\,s\in S(\varphi),\, t\in S(\psi)\},
\]
and thus
\begin{eqnarray*}
   L(\varphi\circ\psi)&=& \max\{\ell(s)\,\,|\,\,s\in S(\varphi\circ\psi)\\
   &\leq& \max\{\ell(s)\,\,|\,\,s\in S(\varphi)\circ S(\psi)\}\\
   &=&\max\{\ell(s\circ t)\,\,|\,\, s\in S(\varphi),t\in S(\psi)\}\\
   &=&\max_{s\in S(\varphi)} \ell(s) + \max_{t\in S(\psi)} \ell(t)=L(\varphi)+L(\psi).
\end{eqnarray*}
Let now $s_{max}$ and $t_{max}$ be elements of maximal length in $S(\varphi)$ and $S(\psi)$ respectively.
Clearly, $\langle s_{max}\circ t_{max}|\varphi\circ\psi\rangle=\sum\alpha_s\beta_t$, where the sum is over
all $s\in S(\varphi)$ and $t\in S(\psi)$ such that $s\circ t=s_{max}\circ t_{max}$. But because of the maximum
length property of $s_{max}$ and $t_{max}$, it follows that $\ell(s)=\ell(s_{max})$ and $\ell(t)=\ell(t_{max})$,
and thus $s=s_{max}$ and $t=t_{max}$. Consequently, $\langle s_{max}\circ t_{max}|\varphi\circ\psi\rangle=\alpha_{s_{max}}
\beta_{t_{max}}\neq 0$, and $L(\varphi\circ\psi)\geq L(\varphi)+L(\psi)$.
\qed
We explain the meaning of these results for lossless quantum data compression below after Definition~\ref{DefLosslessCode},
the definition of a lossless 	quantum code.

\section{Lossless Quantum Data Compression}
\label{SecLosslessQDC}
Our aim is to compute the best possible rate for compressing the unknown output of
a single quantum information source, where the source is given by an ensemble
$\mathcal{E}=\{p_i, \ket{\psi_i}\}_i$ of in general non-orthogonal quantum states $\ket{\psi_i}$
with probabilities $p_i>0$.
As motivated in the Introduction, we want to minimize the expected base length of the code,
and we want to use a prefix-free code to allow concatenation of code words in the case
of several sources.

\begin{definition}[Lossless Quantum Code]
\label{DefLosslessCode}
\lineclear
Let $\mathcal{E}=\{p_i, \ket{\psi_i}\}_i$ be an ensemble of quantum states in a Hilbert space,
with $\hr:={\rm span}\{|\psi_1\rangle,\ldots,|\psi_n\rangle\}$.
A lossless code $C$ is an isometric linear map from $\hr$ into a closed prefix-free subspace $\hr'\subset\hr_{\{0,1\}^*}$.

The expected base length of compression of $C$ is
\begin{equation}
   E(L(C(\mathcal{E}))) = \sum_i p_i L(C(\ket{\psi_i})).
\label{EqDefExpLength}
\end{equation}
$C$ is optimal if for any other code $C'$,
\[
   E(L(C(\mathcal{E})) \leq E(L(C'(\mathcal{E}))).
\]
\end{definition}

The expression (\ref{EqDefExpLength}) defines the compression rate of the code as the expected base length
of the encoding of the output of a {\em single} instance of the ensemble. What if we have {\em $n$ copies} $\mathcal{E}^{\otimes n}$
of an ensemble $\mathcal{E}$,
i.e. several output states are produced independently and identically distributed according to $\mathcal{E}$?

Suppose we have two different ensembles $\mathcal{E}=\{p_i,\ket{\psi_i}\}$ and
$\mathcal{F}=\{q_j,\ket{\varphi_j}\}$ which have optimal codes $C_{\mathcal{E}}$
and $C_{\mathcal{F}}$ respectively. As the codes are prefix-free, we may concatenate them to obtain a code $C_{\mathcal{E}}\circ C_{\mathcal{F}}$
for $\mathcal{E}\otimes\mathcal{F}$. Theorem~\ref{TheIsoConcat}
proves that this concatenation can be done unitarily, i.e. can be implemented in principle on a quantum computer,
and Lemma~\ref{LemAdditivity} tells us that the base lengths then just add up. Explicitly,
\begin{eqnarray*}
E(L(C_{\mathcal{E}}\circ C_{\mathcal{F}}))&=&\sum_{ij}p_i q_j L(C_{\mathcal{E}}(\ket{\psi_i})\circ C_{\mathcal{F}}(\ket{\varphi_j}))\\
&=&\sum_{ij} p_i q_j \left(\strut L(C_{\mathcal{E}}(\ket{\psi_i})+L(C_{\mathcal{F}}(\ket{\varphi_i}))\right)\\
&=&E(L(C_{\mathcal{E}}))+E(L(C_{\mathcal{F}})).
\end{eqnarray*}
Thus $C_{\mathcal{E}}\circ C_{\mathcal{F}}$ is a code for $\mathcal{E}\otimes\mathcal{F}$
with the simple property that its rate is just the sum of the rates of the two codes.
However, it is not necessarily optimal any more. In fact, denoting the optimal compression rate of an ensemble $\mathcal{E}$
by $R(\mathcal{E})$, Theorem~\ref{MainTheorem} below will show that if e.g.
\[
   \mathcal{E}:=\left\{\left(\frac 1 3,\frac 1 3,\frac 1 3\right),\left(
   \ket{0},\ket{1},\ket{2}\right)\right\}
\]
where $\ket{0}$, $\ket{1}$ and $\ket{2}$ are three arbitrary orthonormal vectors,
then $R(\mathcal{E})=\frac 5 3$, while $R(\mathcal{E}\otimes\mathcal{E})=\frac{29}9<2R(\mathcal{E})=\frac{30}9$.
Hence concatenation of codes does not always produce optimal codes (although they are typically quite good), and our result will
for example not give a simple expression for the asymptotic rate $\lim_{n\to\infty}\frac 1 n R(\mathcal{E}^{\otimes n})$,
only the upper bound $R(\mathcal{E})$.

Yet, the result is nevertheless useful, in particular if there is only one output of the source,
or if there are several sources $\mathcal{E}_1\otimes\mathcal{E}_2\otimes\ldots\otimes\mathcal{E}_k$ which are
not known in advance to the compressor. Then, the compression can be done sequentially, for one source after the
other, and the code words are concatenated while the rates just add up. As for the compression rate,
we get the useful upper bound $R\leq\sum_{i=1}^k R(\mathcal{E}_i)$ even if there is no translation invariance
in the sequence of sources.

This subadditivity property of the optimal rate also shows that in the case of $n$ copies of one source,
block coding with concatenation will produce the optimal asymptotic compression rate: write
\[
   \mathcal{E}^{\otimes n}=\mathcal{E}^{\otimes n_1}\otimes \mathcal{E}^{\otimes n_2}\otimes\ldots\otimes \mathcal{E}^{\otimes n_k}
\]
with $\sum_{i=1}^k n_k=n$ such that the sequence $(n_k)_{k\in\N}$ is increasing. Then, use the optimal code $C_{n_k}$ for each
block $\mathcal{E}^{\otimes n_k}$ separately, and concatenate the codes to get a code for $\mathcal{E}^{\otimes n}$.
The corresponding compression rate will be asymptotically optimal.

To state the optimal compression rate for single sources, we introduce the notion of {\em monotone entropy} and of a {\em sequential projection} of
some ensemble $\mathcal{E}=\{p_i,|\psi_i\rangle\}$.
\begin{definition}[Monotone Entropy]
\lineclear
Let $p=(p_1,p_2,\ldots,p_n)$ be a probability vector. Then, we define the {\em monotone entropy} $H_{mon}(p)$ as
\begin{eqnarray*}
   H_{mon}(p)&:=&\min\left\{\sum_{i=1}^n p_i \ell_i \,\,\left\vert \,\, \sum_{i=1}^n 2^{-\ell_i}\leq 1,\right.\right.\\
   &&\left.\enspace\qquad\underbrace{\ell_1\leq \ell_2\leq\ldots\leq \ell_n}_{(*)},\enspace \ell_i\in\N_0
   \right\}.
\end{eqnarray*}
\end{definition}
Note that the Kraft inequality on the right-hand side implies that the values $\{\ell_i\}_i$ are code word
lengths of a prefix code.

Suppose we removed $(*)$ from the definition. This would mean that we look for the smallest possible rate of
any prefix code for the given probability distribution $p$. As is well-known, this best rate is given by
the Shannon entropy $H(p)$; thus, we would get back (up to possibly one bit) Shannon entropy. This implies
\begin{equation}
   H_{mon}(p)\geq H(p),
   \label{HmonLowerBound}
\end{equation}
and justifies that we call $H_{mon}$ an {\em entropy}. Note that $H_{mon}$ changes if we permute the entries of $p$
(while Shannon entropy stays constant). If the elements of $p$ are in decreasing order, then monotone entropy
equals Shannon entropy up to possibly one bit:
\begin{equation}
   p_1\geq p_2\geq \ldots \geq p_n\enspace\Rightarrow \enspace H(p)\leq H_{mon}(p)\leq H(p)+1.
   \label{HmonOrdered}
\end{equation}
This is easily proved by inserting $\ell_i:=\lceil -\log p_i\rceil$.
On the other hand, if we set $\ell_i:=\lceil \log n\rceil$ for every $i$, we get the universal upper bound
\begin{equation}
   H_{mon}(p)\leq \lceil \log n\rceil,
   \label{HmonUpperBound}
\end{equation}
if $n$ denotes the number of elements in $p$.

Now we explain the notion of a {\em sequential projection}.
It is a certain probability distribution which is constructed from $\mathcal{E}$
in a sequential manner.
\begin{definition}[Sequential Projection]
\label{DefSequentialProjection}
Let $\mathcal{E}=\{(p_i,\ket{\psi_i})\}_{i=1}^n$ be an ensemble of quantum states.
A {\em sequential projection} $p'=(p_1',p_2',\ldots,p_k')$ is any probability distribution
which can be constructed by the following algorithm:
\begin{itemize}
\item Choose an arbitary integer $i_1\in\{1,\ldots,n\}$. Then, add up all the probabilities $p_j$ that correspond
to vectors $\ket{\psi_j}$ which are linearly dependent on (parallel to) $\ket{\psi_{i_1}}$ to get the value $p_1'$, i.e.
\[
   I_1:=\left\{j\in\{1,\ldots,n\}\,\,|\,\,\ket{\psi_j}\in{\rm span}\{\ket{\psi_{i_1}}\}\right\}
\]
(in particular, $i_1\in J$), and $p_1':=\sum_{j\in I_1} p_j$.
\item Choose an arbitrary remaining integer $i_2\in\{1,\ldots,n\}\setminus I_1$. Add up all the probabilities $p_j$
that correspond to vectors $\ket{\psi_j}$ which are linearly dependent on $\ket{\psi_{i_2}}$ and the previously chosen vectors in $I_1$
to get the value $p_2'$, i.e.
\[
   I_2:=\left\{j\in\{1,\ldots,n\}\setminus I_1\,\,|\,\,\ket{\psi_j}\in
   {\rm span}\left(\{\ket{\psi_{i_2}}\}\cup I_1\right)\right\}
\]
and $p_2':=\sum_{j\in I_2} p_j$.
\item Choose an arbitrary remaining integer $i_3\in\{1,\ldots,n\}\setminus \left(I_1\cup I_2\right)$. Add up all the probabilities $p_j$
that correspond to vectors $\ket{\psi_j}$ which are linearly dependent on $\ket{\psi_{i_3}}$ and the previously chosen vectors
to get the value $p_3'$, i.e.
\begin{eqnarray*}
   I_3&:=&\left\{j\in\{1,\ldots,n\}\setminus \left(I_1\cup I_2\right)\,\,|\right.\\
   &&\left.\qquad\qquad\qquad\qquad\,\,\ket{\psi_j}\in{\rm span}\left(\{\ket{\psi_{i_3}}\}
   \cup I_1\cup I_2\right)\right\}
\end{eqnarray*}
and $p_3':=\sum_{j\in I_3} p_j$.
\item $\ldots$
\item Iterate these steps until there are no remaining vectors in the ensemble.
\end{itemize}
\end{definition}
As an example of a sequential projection, consider the states from an ensemble $\{p_i,\ket{\psi_i}\}_{i=1}^4$
\begin{eqnarray*}
\ket{\psi_1}&=&\ket{0},\quad
\ket{\psi_2}=\ket{+} = \frac{\ket{0}+\ket{1}}{\sqrt{2}},\\
\ket{\psi_3}&=&\ket{1},\quad
\ket{\psi_4}=\ket{2},
\end{eqnarray*}
where $\ket{0}$, $\ket{1}$ and $\ket{2}$ denote orthonormal basis vectors from an arbitrary Hilbert space.
Then, applying the definition above, and noting that $\ket{\psi_3}$ is in the span of $\ket{\psi_1}$ and $\ket{\psi_2}$,
we get one possible sequential projection as
\[
   p_1'=p_1,\quad p_2'=p_2+p_3,\quad p_3'=p_4,
\]
where we have chosen $i_1=1$, $i_2=2$ and $i_3=4$. Other choices of indices yield different sequential projections. That is,
to every ensemble $\mathcal{E}$, there are several possible sequential projections of $\mathcal{E}$. By combinatorics, the number of
sequential projections to an ensemble $\mathcal{E}$ of $n$ elements is upper-bounded by $n!$. Each sequential projection is
a probability vector with $\dim \mathcal{E}$ elements.

To get an idea how sequential projections are related to base length compression, suppose a code $Q$ compresses the state $\ket{0}$
with base length $l_1$ and the state $\ket{+}$ with base length $l_2\geq l_1$, then, since $\ket{1}$ is on the span of $\ket{0}$ and $\ket{+}$,
$\ket{1}$ will typically also be compressed to $l_2$.
Suppose $\ket{2}$ is compressed to length $l_3$ (which can safely be achieved if $l_3\geq l_2$), then
the compression rate of $\mathcal{E}$ is
\[
   E(L(Q(\mathcal{E}))) = p_1 l_1 + (p_2+p_3)l_2 + p_4 l_3 = p_1' l_1 + p_2' l_2 + p_3' l_3.
\]

We can now state the optimal rate of compression in terms
of monotone entropy and sequential projection, which can
both be calculated combinatorially.
\begin{theorem}[Optimal Compression Rate]
\label{MainTheorem}
\lineclear
Let $\mathcal{E}=\{p_i,|\psi_i\rangle\}$ be an ensemble of quantum states in some Hilbert space. Then the optimal base length
lossless quantum prefix compression code $C$ can be constructed such that it maps into a Hilbert space $\hr'$ which is spanned by an orthonormal
basis of length eigenvectors. The rate $R$ of this optimal code is given by
\[
   R=\min\{H_{mon}(p')\,\,|\,\, p'\mbox{ is a sequential projection of }\mathcal{E}\}.
\]
In particular, if the vectors $\{|\psi_i\rangle\}_i$ are linearly independent, then $H(p)\leq R\leq H(p)+1$,
i.e. the rate is essentially given by the Shannon entropy of $\mathcal{E}$'s probability distribution. In any case,
we have the upper bound $R\leq H(p)+1$.
\end{theorem}
Before we give a proof, we illustrate the theorem with one example. Suppose our ensemble consists of
eight states $\ket{\psi_1},\ket{\psi_2},\ldots,\ket{\psi_8}$ from some Hilbert space, each with probability
$p_1=p_2=\ldots=p_8=\frac 1 8$, such that the span
of those eight states has dimension four. Furthermore, suppose that any four of those states are linearly independent.

Our theorem tells us that we can compress the ensemble at least as good as $R\leq H(p)+1=H\left(\frac 1 8,\frac 1 8,
\ldots,\frac 1 8\right)+1=4$, but we can do better than that. To compute the optimal compression rate,
we have to look at all possible sequential projections.

We construct a sequential projection $p'$: first, we arbitrarily choose one of the vectors, say, $\ket{\psi_1}$.
As there is no other vector which is linearly dependent on (parallel to) $\ket{\psi_1}$, the first entry to $p'$
is $p'_1:=p_1=\frac 1 8$.

As the second step, we choose one of the remaining vectors, say, $\ket{\psi_2}$. We have to check if there are
any remaining vectors that are in the span of $\ket{\psi_1}$ and $\ket{\psi_2}$ (i.e. linearly dependent on those two),
which is by assumption not the case. Thus, we get $p'_2:=p_2=\frac 1 8$.

We go on by choosing the next vector arbitrarily, say, $\ket{\psi_3}$. As there are no remaining vectors in the
span of $\ket{\psi_1}$, $\ket{\psi_2}$ and $\ket{\psi_3}$, we also get $p'_3:=p_3=\frac 1 8$.

Then we select another remaining vector, say, $\ket{\psi_4}$. But now, all the remaining vector $\ket{\psi_5}$,
$\ket{\psi_6}$, $\ket{\psi_7}$ and $\ket{\psi_8}$, are in the linear span of $\ket{\psi_1}$, $\ket{\psi_2}$,
$\ket{\psi_3}$ and $\ket{\psi_4}$. Thus, we have to add the corresponding probabilities to get
$p'_4:=p_4+p_5+p_6+p_7+p_8=\frac 5 8$. Thus,
\[
   p'=\left(\frac 1 8,\frac 1 8,\frac 1 8,\frac 5 8\right).
\]

In this example, repeating the process with different choices of vectors will always result in the
same probability distribution $p'$. Thus, in this case, there is only one possible sequential projection
of $\mathcal{E}$ which is given above. We get the rate $R$ by computing $R=H_{mon}(p')$. First, we know
from (\ref{HmonUpperBound}) that $H_{mon}(p')\leq \lceil \log 4 \rceil=2$. In fact, with the help of
a little computer program, it is easy to see that the minimum in the definition of $H_{mon}$ is indeed
attained at this value, that is,
\[
   R=H_{mon}(p')=2.
\]
Now we prove this theorem.

{\bf Proof.} The proof consists of two parts: first, we show that a rate of $R$ is achievable, then we show that
this rate is optimal. We shall denote our ensemble by
$\mathcal{E}=\{\underbrace{p(|\psi_i\rangle)}_{=:p_i},|\psi_i\rangle\}_{i=1}^n$, and we write
$\hr:={\rm span}\{|\psi_1\rangle,\ldots,|\psi_n\rangle\}$. For sequential projections, we use
the nomenclature from Definition~\ref{DefSequentialProjection}.

To see the achievability, let $p'=(p'_1,\ldots,p_d')$ be an arbitrary sequential projection of $\mathcal{E}$.
Let $(c_1,\ldots,c_d)\subset\{0,1\}^*$ be a prefix code with code word lengths $\ell_i:=\ell(c_i)$
which are minimizers in the definition of $H_{mon}(p')$ (such a code exists due to the Kraft inequality).
Let $\hr':={\rm span}\{|c_1\rangle,\ldots,|c_d\rangle\}\subset\hr_{\{0,1\}^*}$. We will now construct a code
(a linear isometric map) $C:\hr\to\hr'$.
For $i\in\{1,\ldots,d\}$,
let $\Psi_i$ be the set of vectors from $\mathcal{E}$ that have been chosen in step $i$ of the construction of $p'$,
such that $\Psi_i\cap \Psi_j=0$ for $i\neq j$, $\cup_{i=1}^d \Psi_i=\{|\psi_1\rangle,\ldots,|\psi_n\rangle\}$,
and $p'_i=\sum_{|\psi\rangle\in\Psi_i} p(|\psi\rangle)$.

We start by specifying the action of $C$ on the vectors of $\Psi_1$. All the vectors in $\Psi_1$ are equal up to some
phase factor, i.e. they are equal to $e^{i\theta} |\psi\rangle$, where $|\psi\rangle\in\Psi_1$. We set
\begin{equation}
    C|\psi\rangle:=|c_1\rangle,
    \label{EqPsi1}
\end{equation}
then $L(C(|\psi\rangle))=\ell(c_1)$ for every $|\psi\rangle\in\Psi_1$. Since $\dim {\rm span}\left(
\Psi_1\cup\Psi_2\right)=2$, we can construct $C$ such that
\begin{equation}
   C\left( {\rm span}\left(\Psi_1\cup\Psi_2\right)\right)={\rm span}\{|c_1\rangle,|c_2\rangle\}
   \label{EqPsi2}
\end{equation}
isometrically, while still respecting (\ref{EqPsi1}). Consequently, $L(C(|\psi\rangle))=\max\{\ell(c_1),\ell(c_2)\}
=\ell(c_2)$ for
every $|\psi\rangle\in\Psi_2$ (here we use the monotonicity property $\ell(c_1)\leq \ell(c_2)\leq\ldots$).
The next step is to demand
\[
   C\left({\rm span}\left(\Psi_1\cup\Psi_2\cup\Psi_3\right)\right)={\rm span}\{|c_1\rangle,|c_2\rangle,|c_3\rangle\}
\]
isometrically, while respecting (\ref{EqPsi1}) and (\ref{EqPsi2}). Iterating this process, we obtain a code $C$
in the sense of Definition~\ref{DefLosslessCode}. The expected base length compression rate is
\begin{eqnarray*}
r&:=& \sum_{j=1}^n p_j L(C(|\psi_j\rangle))=\sum_{i=1}^d \left(\sum_{|\psi\rangle\in\Psi_i}p(|\psi\rangle)\right)\ell(c_i)\\
&=&\sum_{i=1}^d p'_i \ell_i=H_{mon}(p').
\end{eqnarray*}

Next, we show that this code is optimal, i.e. no lossless code can beat monotone entropy. Thus, let $C:\hr\to\hr'$ be a code in
the sense of Definition~\ref{DefLosslessCode}, and let $r(C):=\sum_i p_i L(C(|\psi_i\rangle))$ be the corresponding
compression rate. We may assume that the vectors $\ket{\psi_i}$ are ordered such that $i<j\Rightarrow L(C(\ket{\psi_i}))
\leq L(C(\ket{\psi_j}))$.

We will now construct a sequential projection $p'$ which corresponds to this code $C$.
Let $i_1:=1$, and $l_1:=L(C(\ket{\psi_{i_1}}))$. Suppose
$\ket{\psi_j}\in I_1$, then $\ket{\psi_j}$ is linearly dependent on $\ket{\psi_{i_1}}$. Since $C$ is isometric,
$C(\ket{\psi_j})$ must be linearly dependent on $C(\ket{\psi_{i_1}})$ as well, and so $L(C(\ket{\psi_j}))=l_1$. So
\[
   \sum_{j\in I_1} p_j L(C(\ket{\psi_j}))=\left(\sum_{j\in I_1} p_j\right) l_1=p_1' l_1.
\]
Then, let $i_2$ be the smallest natural number which is not in $I_1$,
and let $l_2:=L(C(\ket{\psi_{i_2}}))$. If $\ket{\psi_j}\in I_2$, then $\ket{\psi_j}$ is in the
linear span of $I_1$ and $\ket{\psi_{i_2}}$. Since $C$ is isometric, we can again conclude that $L(C(\ket{\psi_j}))\leq l_2$.
But if we had $L(C(\ket{\psi_j}))< l_2=L(C(\ket{\psi_{i_2}}))$, then it would follows that $j<i_2$ which is impossible.
Hence $L(C(\ket{\psi_j}))=l_2$, and
\[
   \sum_{j\in I_2} p_j L(C(\ket{\psi_j}))=\left(\sum_{j\in I_2} p_j\right) l_2=p_2' l_2.
\]
We iterate this procedure until all the vectors from the ensemble have been used. Since the vectors
$\ket{\psi_i}$ are ordered according to their lengths, we have $l_1\leq l_2\leq\ l_3\leq \ldots$ and so on.
Moreover, these code word lengths satisfy the Kraft inequality. To see this, note that the vectors $\ket{\psi_{i_k}}$
are linearly independent and span the Hilbert space $\hr'$. Let $\{\ket{\varphi_k}\}_k$ be the orthonormal basis
of $\hr'$ which is generated by the Gram-Schmidt orthonormalization process from the basis $\{\ket{\psi_{i_k}}\}_k$.
It follows that
\[
   L(\ket{\varphi_k})\leq\max_{k'\leq k} L(C(\ket{\psi_{i_{k'}}}))=L(C(\ket{\psi_{i_k}}))=l_k.
\]
Since $\hr'$ is a prefix Hilbert space, the quantum Kraft inequality from Lemma~\ref{LemKraft} yields
\begin{equation}
   \sum_k 2^{-l_k}\leq \sum_k 2^{-L(\ket{\varphi_k})} \leq 1.
   \label{EqPhiK}
\end{equation}
Moreover, $p'=(p_1',p_2',\ldots)$ is by construction a sequential projection. Hence
\[
   r(C)=\sum_{j=1}^n p_j L(C(\ket{\psi_j}))=\sum_k p_k' l_k \geq H_{mon}(p'),
\]
which concludes the optimality part of the proof. An easy additional argument shows that the optimal code
Hilbert space $\hr'$ may always be chosen to be spanned by an orthonormal basis of length eigenstates:
Due to (\ref{EqPhiK}), there is a classical prefix-free code $\{c_k\}_k$ with $\ell(c_k)=L(\ket{\varphi_k})$.
Let $\hr'':={\rm span}_k \ket{c_k}$, then $\hr''$ is prefix-free. Let $U\ket{\varphi_k}:=\ket{c_k}$, then
$U$ maps $\hr'$ unitarily onto $\hr''$. Hence, the composition $U\circ C$ is a lossless quantum code. Suppose
$j\in I_k$, then $\ket{\psi_j}\in{\rm span}_{k'\leq k} \ket{\psi_{i_{k'}}}$, hence
\[
   U\circ C(\ket{\psi_j})\in{\rm span}_{k'\leq k} U\circ C(\ket{\psi_{i_{k'}}}) ={\rm span}_{k'\leq k} U\ket{\varphi_{k'}},
\]
and so
\begin{eqnarray*}
   L(U\circ C(\ket{\psi_j}))&\leq&\max_{k'\leq k} L(U\ket{\varphi_{k'}})=\max_{k'\leq k} \ell(c_k)\\
   &=&\max_{k'\leq k} L(\ket{\varphi_{k'}})\leq \max_{k'\leq k} L(\ket{\psi_{i_{k'}}})\\
   &=&\max_{k'\leq k} l_{k'} = l_k = L(C(\ket{\psi_j})).
\end{eqnarray*}
Thus, $r(U\circ C)\leq r(C)$, i.e. $U\circ C$ compresses at least as good as $C$.

In the special case that all the vectors $\ket{\psi_i}$ are linearly independent, the sequential projections of $\mathcal{E}$
are exactly the permutations of the probability distribution $p$. Using (\ref{HmonLowerBound}) and (\ref{HmonOrdered}),
we thus get
\begin{eqnarray*}
   R&=&\min_{p'} H_{mon}(p')=\min_{\sigma\mbox{ permutation}} H_{mon}(\sigma(p))\\
   &\in& [H(p),H(p)+1].
\end{eqnarray*}
It remains to prove that the optimal rate is always bounded above by $H(p)+1$. For this purpose,
rearrange the vectors $\ket{\psi_i}$ in decreasing order such that $p_1\geq p_2\geq\ldots\geq p_n$.
Let $p'$ be the sequential projection which is constructed by getting through the list of $\ket{\psi_i}$'s in
that order. As before, denote by $\Psi_j$ the set of $\ket{\psi_i}$'s that
have been collected in step $j$ of the construction of $p'$. Let
\[
   \ell_i:=\lceil -\log \max_{\ket\psi\in \Psi_i} p(\ket\psi)\rceil.
\]
By construction, $\ell_1\leq\ell_2\leq\ldots\leq \ell_d$, and the Kraft inequality holds for the $\ell_i$. Thus,
\begin{eqnarray*}
   R&\leq&H_{mon}(p')\leq \sum_{i=1}^d p'_i \ell_i\\
   &=&\sum_{i=1}^d \left(\sum_{\ket\psi\in \Psi_i}p(\ket\psi)\right)
   \lceil -\log \max_{\ket\psi\in \Psi_i} p(\ket\psi)\rceil\\
   &\leq& \sum_{i=1}^d \sum_{\ket\psi\in \Psi_i} p(\ket\psi)\lceil -\log p(\ket\psi)\rceil\leq H(p)+1.
\end{eqnarray*}
This proves the statement of the theorem.
\qed

\section{Conclusions}
\label{SecConclusions}
We have given a method for lossless compression of unknown outputs of single quantum information
sources which minimizes the code's expected base length, and we have calculated the corresponding optimal
compression rate (Theorem~\ref{MainTheorem}).
Moreover, we have explained a simple model of an always-open channel which admits the lossless
transmission of the indeterminate-length code words, and we have explained that compression
can reduce transmission errors for those channels.

As our approach quantifies the rate in terms of the base length,
it complements work by Schumacher and Westmoreland~\cite{sch01}
who have given the optimal rate for average length compression.
Furthermore, we have demonstrated how to apply the theory of prefix-free subspaces
to quantum information. In short, prefix-free quantum strings allow sequential compression
in the case of several quantum information sources by concatenating the corresponding
code words. The concatenation can be accomplished physically (Theorem~\ref{TheIsoConcat}), even in the case
of prefix-free subspaces which are more general then in Schumacher and Westmoreland's sense (cf. Example~\ref{ExStrangeHR}).

At this point, it remains open if there is a simple formula for the optimal asymptotic compression
rate $\lim_{n\to\infty} \frac 1 n R(\mathcal{E}^{\otimes n})$ in the case of $n$ copies of
a single source $\mathcal{E}$, apart from the upper bound $R(\mathcal{E})$.
Also, it would be nice to have an example of a physical situation where base length compression is better
suited to reduce transmission errors for channels than average length compression (cf. Section~\ref{SecCommunicationModel}).
Even though the optimal asymptotic compression rate is not given in this paper, the result is optimal
for the case of a sequence of several sources $\mathcal{E}_1\otimes\mathcal{E}_2\otimes\ldots\otimes\mathcal{E}_k$
which are not known in advance and have to be compressed sequentially.

Many open questions in quantum information theory, such as entanglement catalysis~\cite{plenio},
are phrased as ``How can this state be transformed into that state exactly and without error subject to these conditions?".
Perhaps lossless quantum base length compression can be applied to some of these questions.
Bostr\"{o}m and Felbinger~\cite{bos02}
stated that lossless quantum compression may also have applications in cryptography. Perhaps it can be used to minimise
the probability that an eavesdropper discovers any information at all, rather than the average information that the eavesdropper
discovers~\cite{nielsen}.

Another possible connection to existing work is in the definition of quantum Kolmogorov complexity by
Berthiaume et al.~\cite{ber01}. They define the complexity of a quantum bit string as the length of
its shortest determinate-length description. Therefore we might expect there to be a close correlation
between this kind of complexity and the rate of compression described in this paper, in the same way that there
is a close correlation between classical Kolmogorov complexity and Shannon entropy.

Apart from possible applications, one purpose of this paper was to show that prefix-free quantum bit strings
are a mathematical structure with nice properties that can be useful in quantum information theory. It might be
interesting to study them in more detail, in particular in connection to possible quantum versions of algorithmic
probability.

\section*{Acknowledgments}
We thank Nihat Ay, Sougato Bose, Arleta Szko\l a, Andreas Winter, and the
members of the Information Theory and Cognitive Systems Group at MPI Leipzig for interesting and helpful discussions.

The work of C. Rogers and R. Nagarajan was partially supported by UK
EPSRC grants (GR/S34090 and EP/E00623X/1) and the EU Sixth Framework
Programme (Project SecoQC: Development of a Global Network for Secure
Communication based on Quantum Cryptography).

\bibliography{losslessqpcompression}

\end{document}